\documentstyle[12pt]{article}

\input tcilatex

\begin{document}

\baselineskip=22truebp

\centerline {\bf INTRUDER STATES AND THEIR LOCAL EFFECT }

\centerline {\bf ON SPECTRAL STATISTICS} \medskip
\centerline{
J. Flores $ ^{1,2 } $, H. Hern\'andez-Salda\~na $ ^2 $, F. Leyvraz $ ^2 $
and T.H. Seligman $ ^{2,3} $}

\centerline{Centro Internacional de Ciencias}
\centerline{Ciudad Universitaria, Chamilpa, Cuernavaca, M\'exico}

\bigskip\bigskip\bigskip\bigskip

\centerline{Abstract} \medskip

The effect on spectral statistics and on the revival probability of intruder
states in a random background is analysed numerically and with perturbative
methods. For random coupling the intruder does not affect the GOE spectral
statistics of the background significantly, while a constant coupling causes
very strong correlations at short range with a fourth power dependence of
the spectral two-point function at the origin.The revival probability is
significantly depressed for constant coupling as compared to random coupling.

\vfill
{$^1 $ Centro Universitario de Comunicaci\'on de la Ciencia}

{$^2 $ Instituto de F\'\i sica}

{$^3 $ Instituto de Matem\'aticas}

{University of M\'exico (UNAM) M\'exico}

\eject
{\bf I. Introduction }\medskip

About a quarter of a century ago the effect of an intruder state on the
spectral statistics of a Gaussian orthogonal ensemble (GOE) was discussed
[1] in the context of isobaric analogue states in nuclei. The limited
numerical possibilities revealed a very small effect in short range
correlations. Recently the problem of intruder states has regained interest
in the context of the analysis of revival times [2] and of state sensitivity
[3] of quantum chaotic systems. The original model [1] proposes a random
coupling of the intruder with the background, quite similar to the one
proposed in [3]. Ref. [2], on the other hand, proposes a constant coupling
and finds an anomalous behaviour in the revival probability.

In the present paper we shall discuss differences between the two models in
terms of spectral statistics. The random coupling case is revisited and with
better statistics the deviations from the statistics of the background seems
even smaller. Constant coupling on the other hand introduces a quartic
repulsion between levels, which leads for the particular case of the
intermediate coupling chosen in [2] to a nearest-neighbour spacing
distribution that is quite similar to the one of a Gaussian simplectic
ensemble (GSE); at intermediate ranges we find a stiffness of spectral
fluctuations that exceeds even the one of a GSE. At long ranges the entire
effect disappears and we revert to the GOE behaviour.

Next we analyze the revival probability. On average it is significantly
larger for random coupling than for constant coupling. This is readily
understood in terms of the size of the intensity fluctuations. Furthermore,
for the constant coupling case the revival probability shows a peak, that is
related to the anomalous spectral statistics mentioned above. \bigskip

{\bf II. The models} \medskip

The basic question we wish to address refers to the effect of an intruder
state immersed into a "chaotic" background of states. This background,
according to a common assumption, can be modeled by a GOE for a time
reversal invariant system. The open question refers to the type of coupling
between the background and the intruder. In the early paper [1] the natural
assumption was made, that the coupling matrix elements should be drawn
randomly from a Gaussian distribution, but that there should be a variable
parameter to change the strength of the coupling. We shall henceforth refer
to this case as model I. The model used in [3] is quite similar.

Model II proposed in [2] is even simpler as it chooses the same coupling
matrix element between the intruder and all background states in the basis
where the background Hamiltonian is diagonal.

We specify our models as follows: Let $H_{-}$ be a $n\times n$ GOE matrix.
Diagonalyse $H_{-}$ and subject the spectrum, centred at zero, to the usual
unfolding procedure [4] normalising the average density to one. We now add
the intruder state in the centre of the spectrum by adding a row and a
column to the matrix. We obtain
$$
H=\left( \matrix{
0          & {\bf v } \cr
{\bf v }^\dagger & H _-   \cr}\right) \eqno (2.1)
$$
where ${\bf v}=(v_1,v_2...v_n)$ gives the coupling between the intruder and
the background states. The two models differ in the choice of the coupling
matrix elements $v_i$.

As mentioned above, model I requires the coupling matrix elements to be
chosen from a Gaussian distribution with width $v=\sqrt{{\overline{v_{i^2}}}}
$. Here $v$ indicates the coupling strength. Actually our model does not
coincide exactly with the one discussed in [1] because there the intruder is
placed in the original matrix. The diagonalisation of the background will
not affect the coupling because an orthogonal transformation will leave ( in
the limit of large $n$) an ensemble of vectors of Gaussian random variables
invariant. The unfolding, on the other hand, does affect the matrix, but we
expect only edge effects, which should vanish at large $n$.

Model II simply chooses $v={\rm constant}$. At $v=2.08$ we are exactly in
the case of [2].

\bigskip

{\bf III. Spectral statistics: numerical analysis} \medskip

In order to obtain information about spectral statistics we have to
diagonalise $H$ and again unfold the resulting spectra. The calculations
were performed with ensembles of $1000$ matrices of dimension $300\times 300$
and their reliability checked by several calculations with ensembles of $%
800\times 800$ matrices. After the diagonalisation of the background states
the fifty lowest and highest states were omitted to avoid finite size
effects.

This was done for both models and various values of $v $. At small $v $ we
are in a perturbative regime, for $v = 1 $ we are in a transition zone which
in model I coincides exactly with the GOE.

For $v =2.08 $ we shall find a spreading width $\Gamma = 27 $ measured in
terms of the average level spacing normalised to one. We show the
intensities of the intruder in the eigenstates in Fig 1.a for model I and in
Fig 1b for model II. The results are given in both cases for an ensemble of
10 matrices. We see that the results fluctuate widely, and more so for model
I than for model II. We may quantify this by calculating the inverse
participation ratio $P^{{\rm I}} = \sum_{i= 1} ^{201} I_i^2 $. Here $I _i =
|a_i|^2 $ are the intensities of the intruder in the eigenstates as shown in
Fig. 1 obtained from the amplitudes $a_i $ that result from the
diagonalisation. We find for models I and II respectively $P^{{\rm I}}_I =
0.036 $ and $P^{{\rm I}}_{II} = 0.019 $, in agreement with the impression we
obtain from the figures.

To obtain a spreading width we smooth the data by performing an ensemble
average and plotting the total intensities of the intruder in states lying
in a unit energy interval. The result of this procedure is shown in Fig. 2
for Model II. The fit with a Lorentzian shows excellent agreement. For model
I a similar result is obtained, but not shown.

At some larger value of $\Gamma $, that depends on the size of the matrix,
the spreading width will reach the dimension of the matrix and edge effects
will become important. For very large and very small couplings we shall
present some perturbation results in the next section, that will essentially
explain our numerical findings.

For model I the nearest-neighbour spacing distribution remains very close to
the one of GOE in accordance with the results reported in [1]. Model II on
the other hand shows a rather surprising development from a distribution
apparently rather similar to the one of GOE for $v=1$ to one resembling a
GSE for $v=5$ as we can appreciate in Fig 3. Fig. 4 shows the same
distribution for $v=1$ and $v=2.08$, but restricted to 10 and 30 spacings
respectively near the centre of the spectrum, which keeps us inside the
corresponding spreading width (see Fig. 2). We now find the GSE type
behaviour fully developed in both cases. Note that the agreement with the
GSE is not very good so we must be careful not to jump to conclusions. On
the other hand the disagreement with a GOE is absolutely convincing.

We next want to check the long range spectral correlations. Here we make use
of a well known fact: As we add a single row and column to a matrix the
eigenvalues of the new matrix are intertwined between the ones of the old
one, as displayed in Fig. 5. Therefore, in both models the long range
behaviour must approach that of a GOE. Actually we expect to find no
significant deviation beyond three average level spacings.

As a consequence of the above exact result we may limit our attention to the
intermediate region. For this region a study of the Fourier transform is
most profitable. Fig. 6 shows such transforms of the unfolded spectra for
the same values of $v $ and the same energy intervals we used for figure 4.
Now we restrict our attention to correlations inside the spreading width
with the same cutoffs as above. Again we see a surprising resemblance with
the GSE. Indeed we find a very marked peak at the value 1 of the transformed
variable which we shall as usual call time. For shorter times though the
correlation hole is even stronger than for a GSE thus confirming our guess
that the coincidence with the properties of this ensemble is limited.

The number variance shown in Fig. 7 for the strong coupling $v = 5 $
displays the expected features. At long range it is parallel to the GOE
curve. At intermediate range we clearly see the transitional behaviour,
which we might have expected. At short range it follows the GSE curve [5]
rather well though the first dip is lower than for the GSE.

\bigskip

{\bf IV. Spectral statistics: Perturbation theory and the quartic level
repulsion }\medskip

{}From many studies of spectra in the transition region [6,7] we know that the
power law behaviour of the spectral two-point function at the origin is most
sensitive to the characteristics of a perturbation. Thus, a GOE perturbed by
the slightest GUE, changes its behaviour immediately [6], though the effect
will be only at very short range. The same was proven for an arbitrary
symmetric or hermitian matrix perturbed by a GOE or a GUE, respectively [7].

To understand the level repulsion in model II, we shall present an argument
that consists of three steps: First we shall show that triplets ( three very
close lying states) in the background necessarily lead at least to doublets
(two very close lying states) in the perturbed system. Second we shall show
that existing doublets are quite unstable and have a small chance of
surviving in model I and none at all in model II. Finally we shall see that
new doublets will only appear in model I. Thus in model II only such
doublets will exist as were generated by a triplet in the background.

The first point is quite simple to show. It again rests upon the fact, used
in the previous section, that by adding one row and one column to a matrix
the new eigenvalues must lie between the ones of the original matrix. If a
triplet exists in the original spectrum one eigenvalue is trapped between
the first two members of the triplet and another between the second and
third members. The two trapped eigenvalues necessarily form a doublet in the
new system. This doublet may or may not form part of a triplet or a quartet
in the enlarged system, but that is not important in the present context.
Recalling the well-known result, that we obtain a GSE by eliminating every
other state from a GOE [8], we find immediately that the probability for a
triplet goes as $x ^4 $ for small level separation $x $. This simultaneously
constitutes an upper bound to the power, as every triplet has to cause a
doublet. Note that this result is non-perturbative.

The second point involves a longer perturbation argument. We consider an
energy doublet $E_1$, $E_2$, coupled weakly to the intruder state. Since we
expect the doublet to couple mainly with itself and, of course with the
intruder, we may disregard all other states. Thus we consider the matrix
$$
\left(
\matrix{
0          & { v_1 }   & { v_2 } \cr
{ v_1 } & { E_1 } & { 0 }  \cr
{ v_2 } & { 0 }  &  { E_2 } \cr}\right) \eqno (4.1)
$$
where we set $0<E_1<E_2$. The secular equation for the eigenvalues of this
matrix can be written as
$$
\lambda ={\frac{v_1^2}{\lambda -E_1}}+{\frac{v_2^2}{\lambda -E_2}}.\eqno %
(4.2)
$$
We are interested in the eigenvalues near the $E_1$-$E_2$ doublet. We
therefore introduce the variables $y=\lambda -E_1$ and $\delta =E_2-E_1$. In
these new variables, the last equation can be rewritten as
$$
y+E_1={\frac{v_1^2}y}+{\frac{v_2^2}{y-\delta }}.\eqno (4.3)
$$
Now comes the crucial approximation: Since the energies $E_1$ and $E_2$ are
a close doublet and since the perturbation due to the intruder is weak, it
is clear that neither the eigenvalue close to $E_1$ nor the one close to $%
E_2 $ will be very far from $E_1$. We can therefore always neglect $y$ with
respect to $E_1$, so that the last equation reduces to
$$
E_1={\frac{v_1^2}y}+{\frac{v_2^2}{y-\delta }},\eqno (4.4)
$$
which yields an equation of second degree, with two solutions $y_1$ and $y_2$%
. For these one readily finds
$$
(y_1-y_2)^2=\left( E_1+v_1^2/E_1-E_2-v_2^2/E_2\right) ^2+{\frac{v_1^2v_2^2}{%
E_1^2}}.\eqno (4.5)
$$
This equation is easy to interpret: In a first approximation, both $E_1$ and
$E_2$ move independently by an amount $v_i^2/E_i$, due to their coupling to
the intruder state. If this motion does not bring them any nearer to each
other than $|v_1v_2/E_1|$, nothing more happens and this result, obtained
from ordinary perturbation theory, remains valid. On the other hand, if the
eigenvalues are closer to each other than $|v_1v_2/E_1|$, then higher orders
of perturbation theory cause the eigenvalues to interact and they wind up at
a distance larger than $|v_1v_2/E_1|$.

Furthermore no doublets can be formed at distances shorter than $%
|v_1v_2/E_1| $. It follows that, for constant $v_i$, no new doublets can be
formed at all. For distributions of the $v_i$ such as the Gaussian, which
allow a finite probability for very small values of the $v_i$, it is
possible to form new doublets from energy levels which were separated before
the coupling. In the above example, this occurs when $v_1\gg v_2$. Then $E_1$
moves a great distance towards $E_2$ and the repulsion does not occur unless
the distance is extremely small.

Thus in model I new pairs are created and again destroyed in terms of
typical avoided crossings and we may reasonably expect to retain the linear
dependence of the GOE two-point function. In model II, on the other hand,
such avoided crossings will not occur and the fourth power limit which we
obtained from the triplet will correspond to the actual distribution.

Note that the same argument can readily be used for an intruder with
constant coupling to a Poisson, GUE or GSE background, to derive a power law
of $x^1,\ x^7 $ and $x^{13}$, respectively. Points two and three of the
above argument transfer immediately and the triplet probabilities for these
ensembles are dominated at short range by the power laws given above.

Our argument is perturbative and thus basically will hold for small
couplings $v$ or for distances outside the spreading width. Figure 8 shows
the level spacing distribution for $v=2.08$ for both models at energies
taken in the wings of the spreading range of the intruder, and we find the
expected $x$ and $x^4$ behaviour for models I and II, respectively. Fig. 9
shows the corresponding results for spacings taken outside the spreading.
Note that in the latter case the expected transition to GOE like behaviour
for model II occurs at very small level separations.

The fact that the figures of the previous section confirm this behaviour
even inside the spreading width, goes beyond the direct scope of our
argument, but is in keeping with the experience discussed above, that
perturbative features at short range tend to extend their range as the
perturbation grows. The deviations from GSE at intermediate and long range
are equally unsurprising as there is no theoretical reason for true GSE
behaviour.

In the limit of very strong interaction for model II the above argument
certainly no longer applies. We actually revert to the GOE except at the
edges of the spectrum. Essentially the only effect of the intruder is to
expel one state at each end of the spectrum, while leaving the remainder
with an unchanged statistic. We may see this by applying perturbation theory
in the opposite sense. That is we consider the singular matrix
$$
H_0=\left( \matrix{
0          & {\bf v } \cr
{\bf v } & {\bf 0}  \cr}\right) \eqno (4.6)
$$
where ${\bf 0}$ is a $n\times n$ matrix of zeros. We choose this matrix as
the unperturbed Hamiltonian and the background Hamiltonian as a
perturbation. The eigenvalues of the unperturbed Hamiltonian are easily
computed to be $E_{\pm }=\pm {\sqrt{n}}v$ and all other eigenvalues are
zero. The eigenfunctions for the non-zero eigenvalues are $\{a_i\}_{\pm
}=\{\pm 1/\sqrt{2},\,1/\sqrt{(}2n),...,1/\sqrt{(}2n)\}$ and the $a_i$ of the
remaining eigenfunctions must fulfill the condition $\sum_{i=1}^na_i=0$ and $%
a_0=0.$ The effect of the perturbation (which is now background) on the
extreme states $\pm $ is well described by first order perturbation theory.
The remaining levels are degenerate and perturbed by a matrix that is very
near to GOE. Therefore, we expect to recover the GOE statisics near the
centre of the spectrum.

\bigskip

{\bf V. The revival probability}

\medskip
Model II was proposed in the context of a revival analysis [2], in which a
numerical calculation of the revival probability turned out to show some
disagreement with a simple theoretical estimate, that used GOE spectral
statistics. We shall proceed to show that such discrepancies are not
resolved if we use the correct spectral statistics. One reason for these
discrepancies will become appearent in terms of the inverse participation
ratio of the intensities.

We recall that the revival probability for the intruder state $\psi (0)$ is
defined as
$$
P(t)=|\langle \psi (0)|\psi (t)\rangle |^2.\eqno (5.1)
$$
If the amplitude of the intruder in the $i$th eigenstate of the full system
is $a_i$, its time evolution is $\psi (t)=\sum_ja_j\exp (iE_j\;t)$. We thus
obtain for the revival
$$
P(t)=\sum_{j,k}|a_j|^2|a_k|^2\exp (i(E_j-E_k)t)=
$$
$$
\sum_i|a_i|^4+\sum_{j<k}|a_j|^2|a_k|^2[\exp (i(E_j-E_k)t)+{\rm c.c.}]\eqno %
(5.2)
$$

For an intruder with Gaussian distributed amplitudes, we recover GOE
properties and the ensemble average for this expresion has been evaluated to
yield%
$$
\langle P(t)\rangle ={\frac 3{{n+2}}}-{\frac{{b(t)}}n}\approx {\ \frac 1n}%
[3-b(\frac t{2\pi })]\eqno (5.3)
$$
for $t>0.$ Here $n$ is the dimension of the matrix and $b(t)$ the usual form
factor [4]. As we have seen that in model I the intruder does not
significantly affect statistics, we may hope that we can extend the
usefulness of  expression (5.3) to this model.

Equation (5.3) is not valid for $t=0$ as $P(0)=1.0$. Indeed the finite
spreading width $\Gamma $ will dominate the short-time behaviour. We shall
heuristically include the effect of the spreading width $\Gamma $ by adding
a short-time term $e^{-\Gamma t}$ and by estimating the effective number of
participating states. For this purpose we assume a smooth distribution of
intensities $I_i$ along the Lorentzian of width $\Gamma $ and calculate the
inverse participation ratio $\sum_i|I_i|^2\approx n_{eff}$. The resulting
effective number of levels is $n_{eff}=\pi \Gamma $ in agreement with ref.
[2]. We thus obtain
$$
\langle P(t)\rangle \approx e^{-\Gamma t}+{\frac 1{{\pi \Gamma }}}[3-b(\frac
t{2\pi })].\eqno
(5.4)
$$
This result clearly contains errors of the order $1/n$ for short times but
we may hope that they are not important as compared to the exponential term.

We performed numerical calculations both for models I and II by
diagonalising the corresponding Hamiltonians for each member of an ensemble
and evaluating $P(t)$ by using eq. (5.2) directly for a range of values of $t
$. The ensemble average was simply performed by averaging the results at
each value of $t$. The results for model II agrees with those obtained in
ref. [2] with other techniques.

We compare in Fig. 10 the theoretical result of eq. (5.4) with the numerical
one for model I and find excellent agreement even for intermediate times,
for which this is not guaranteed by our argument.

The theoretical curve of ref. [2] is also included in the picture and we
find as reported in this reference a very partial agreement with model II.
Note that the theoretical result of ref. [2] corresponds to eq. (5.4) if we
replace the 3 by a 1. This would, in our derivation, correspond to an
absence of fluctuations in Fig. 1b. The $n$-dependence of corrections due to
fluctuations is discussed in ref.[2] but they are not evaluated. We thus
identify two sources of disagreement for model II with theory: First, inside
the spreading width the form factor is not the one of a GOE and second, the
inverse participation ratio must be greater than the inverse of the
effective number of levels, though smaller than for a GOE.

The similarity of our result with that of [2] may tempt us to use a
semi-empirical formula of the same type, where both the inverse
participation ratio and the form factor are taken from the numerical
experiment. Such a formula turns out to be in disagreement with the
numerical result for model II. In particular the very low values of revival
near $t\sim 0.5$ cannot be explained. This indicates that the factorised
structure of the equation is inadequate and is therefore a sign of strong
correlations between energies and amplitudes that go beyond the general
lorentzian shape. Note though that the Fourier transform we found inside the
spreading width gives a qualitative understanding of the maximum in revival
near $t=2\pi .$

\bigskip

{\bf VI. Conclusions}

\medskip
The present paper analysed the effect of an intruder state on a chaotic
background represented by an unfolded spectrum with GOE fluctuations. We
consider both a random Gaussian and a constant coupling to this background.
For the random coupling case we retrieve standard GOE behaviour for spectral
statistics, inverse participation ratio and revival probability of the
intruder, by taking into account the spreading of the intruder in a
straightforward way. For constant coupling on the other hand, we found
inside the spreading width a surprising GSE-like behaviour at short
distances and a GOE behaviour at long range. We could give exact proof for
the latter and a perturbative derivation for the former. As far as the
inverse participation ratio is concerned we find an intermediate value
corresponding to non-vanishing fluctuations of the intensities as seen in
Fig. 1b. A theoretical evaluation is missing.

Concerning the revival probability of the intruder we obtain some
qualitative understanding from the Fourier transforms of the spectrum and
the inverse participation ratio, but the data definitively indicate
important coupling between intensities and energy eigenvalues. A theoretical
understanding is again missing.

It is interesting to note that  the results of sections III and V are
unaffected  if we provide the constant coupling with a random sign. A
possible candidate for an intruder with near constant coupling but random
signs could be an isobaric analogue state in a nucleus, where the $T_{<}$
states are isolated resonances forming the background, while the $T_{>}$
analogue state is the intruder. A separate study will have to show whether
the matrix elements are sufficiently constant to cause the short range
effect discovered in this paper. A further question is whether there are
sufficient data to see this effect in experiment.

\bigskip

{\bf References}

\medskip
\parindent=0pt

[1] Brody T A, Mello P A, Flores J and Bohigas O 1973 {\sl Lettere Nuov. Cim.%
} {\bf 7} 707

[2] Gruver J L, Aliaga J, Cerdeira H A, Mello P A and Prieto A N 1997
Energy-level Statistics and Time Relaxation in Quantum Systems {\sl Phys.
Rev. E.} to appear.

[3] Aiba H, Mizutori S and Susuki T 1997 Fluctuation Properties of Strength
Function Phenomena: A Model Study {\sl Phys. Rev. E.} to appear.

[4] Brody T A, Flores J, French J B, Mello P A, Pandey A and Wong S S M 1981
{\sl Rev. Mod. Phys. }{\bf 53} 385.

[5] {\it Idem}. Eqs. (5.12), (5.13); note a misprint in eq. (5.13), where
the relevant expression should read $\Sigma _4^2(r)=1/2\Sigma
_2^2(2r)+1/(4\pi ^2)[{\rm Si}(2\pi r)]^2$.

[6] French J B, Kota V K\ B, Pandey A and Tomsovic S 1988 {\sl Ann. Phys. NY
}\ {\bf 181} 198

[7] Leyvraz F and Seligman T H 1990 {\sl J. Phys. A} {\bf 23} 1555

[8] Mehta M L and Dyson F J 1963 {\sl J. Math. Phys.} {\bf 4} 713

\newpage\

\bigskip
\centerline{{\bf FIGURE CAPTIONS }} \medskip

Fig. 1: The intensity spectrum $I_i$ of the intruder plotted against the
eigenenergies for an ensemble of ten background matrices with a) Gaussian
random coupling and b) constant coupling to the background. The coupling
strength is in both cases $v=2.08$ \medskip

Fig. 2: Intensities as in the previous figure, but smoothed over an ensemble
of 1000 background matrices showing the average value for energy bins of
size 1 in the normalised spectrum. The results for $v=1.0$ are shown by $%
\times $ and those for $v=2.08$ correspond to $*$. The dotted line gives a
Lorentzian fit with width $\Gamma =6.5$ and the dashed line the same with $%
\Gamma =27.1$ \medskip

Fig. 3: Nearest-neighbour spacing distributions are shown for $v=1.0$
(continuous histogramme) and $v=5$ (dashed histogramme) using an ensemble of
1000 background matrices. The long and short dashed lines show the results
for GOE and GSE respectively. \medskip

Fig. 4: Nearest-neighbour spacing distributions are shown for $v=1.0$
restricted to 10 spacings around zero (dotted histogramme) and for $v=2.08$
(continuous histogramme) for thirty spacings around zero, using an ensemble
of 1000 background matrices. The long and short dashed lines show the
results for GOE and GSE respectively. \medskip

Fig. 5: Graphical solution of the characteristic equation for the
Hamiltonian (2.1). The locations of the poles correspond to the unperturbed
energies and the circles to the energies of the full Hamiltonian. \medskip

Fig. 6: The full line shows the Fourier transform of the spectrum for model
II. In a) for $v=1.0$ restricted to 10 states around zero; in b) for $v=2.07
$ and restricted to 30 states around zero. The transforms for GOE and GSE
are given by long and short dashed lines respectively. \medskip

Fig. 7: The number variance for $v=5.0$ for the whole spectrum ( continuous
line). The results for GOE and GSE are given by long and short dashed lines
respectively. \medskip

Fig. 8: The histogrammes give the nearest-neighbour spacing distributions
for Models I and II taken in the wings of the spreading range. Model I
agrees well with the GOE result (dashed dotted line) and model II with the
GSE (dashed double dotted line). \medskip

Fig. 9: Same as Fig. 8 but taken outside the spreading range. Note the
transition behaviour from GOE to GSE for model II between $x=0.15$ and $%
x=0.25$.\medskip

Fig. 10: The revival of the intruder calculated according to eq. (5.2) for
models I and II. The theory according to eq. (5.4) is given by the short
dashed line and agrees well with model I. The theory of ref. [2] is given as
a long dashed line.

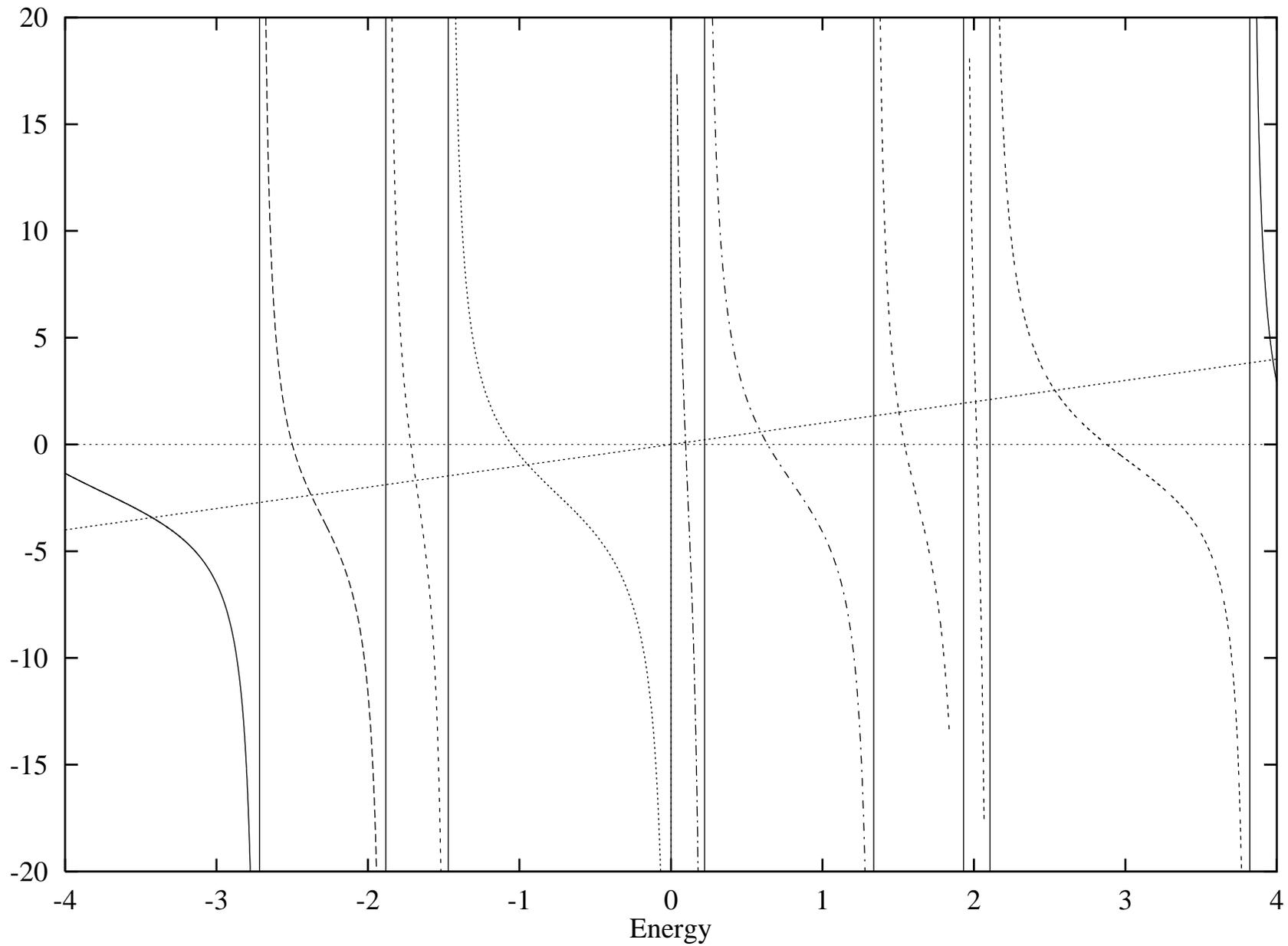
\end{document}